\documentclass[twocolumn,showpacs,preprintnumbers,amsmath,amssymb]{revtex4}
\usepackage{epsfig}
\def\beq{\begin{equation}}
\def\be{\begin{eqnarray}}
\def\eeq{\end{equation}}
\def\ee{\end{eqnarray}}

\def\lsim{\buildrel < \over {_{\sim}}}
\def\gsim{\buildrel > \over {_{\sim}}}
\begin{document}

\title{Estimates of the uncertainties associated with models \\ 
of the nucleon structure functions in the $\Delta$ production 
region}

\author{Omar Benhar}
\author{Davide Meloni}

\affiliation
{
INFN, Sezione di Roma\\
Dipartimento di Fisica, Universit\`a ``La Sapienza'' \\ I-00185 Roma, Italy \\
}

\date{\today}
\begin{abstract}
Theoretical studies of the inclusive electron-nucleus cross section at 
beam energies up to few GeV show that, while the region of the quasi-elastic peak is 
understood at quantitative level, the data in the $\Delta$ production region are
sizably underestimated. We analize the uncertainty associated with the description
of the nucleon structure functions $W_1$ and $W_2$ and its impact on the  
nuclear cross section. The results of our study suggest that the failure to reproduce 
the data is to be mostly ascribed to the poor knowledge of the neutron structure 
functions at low $Q^2$.
\end{abstract}
\maketitle

In view of the rapid development of neutrino physics,
leading to significant improvements in the experimental accuracy,
the treatment of nuclear effects in data analysis is now regarded as one of the
main sources of systematic uncertainty \cite{NUINT01,NUINT04}.

Much of the information needed to understand the nuclear response
to neutrino interactions in the energy range $E_{\nu} =0.5-3$ GeV, 
relevant to many neutrino experiments,
can be extracted from the results of experimental and theoretical studies 
of electron-nucleus scattering (for a recent review see, e.g., Ref. \cite{Benhar06}).
In this kinematical regime, in which  single nucleon knock out is known to be 
the dominant reaction mechanism, both quasi-elastic and inelastic processes, 
leading to the appearance of hadrons other than protons and neutrons, must be
taken into account.

Non relativistic nuclear many-body theory (NMBT) provides a consistent 
framework, suitable to describe electron-nucleus interactions in a variety of 
kinematical 
conditions, ranging from quasi-elastic to deep inelastic scattering 
(see, e.g., Ref. \cite{Benhar93a}). 
In the impulse approximation regime, in which scattering off a nuclear target 
reduces to the incoherent sum of elementary processes involving individual nucleons, 
the basic elements of the NMBT approach are the nucleon spectral function 
$P({\bf p},E)$, yielding 
the energy and momentum distribution of the knocked out nucleon, and the 
electron-nucleon 
cross section, that can be written in terms of the two structure functions $W^N_1$ and 
$W^N_2$ ($N=p,n$).

A study of inclusive electron scattering off oxygen at beam energies between 0.7 and 
1.2 GeV, carried out within NMBT \cite{Benhar05}, has recently shown
that, while the data in the region of the quasielastic peak are accounted for 
with an accuracy better than $\sim$ 10 \%, theory fails to explain the measured cross
sections at larger electron energy loss, where $\Delta$ production 
dominates. Based on the fact that the calculation of the 
cross section at the quasi-elastic and $\Delta$ production peak involves
integrations of $P({\bf p},E)$ extending over regions of the
$({\bf p},E)$ plane which almost exactly overlap one another, 
the authors of Ref. \cite{Benhar05} argued that the disagreement 
between theory and data is unlikely to be imputable to deficiencies of the 
spectral function, and should rather be ascribed to the description of the 
elementary electron-nucleon cross section.

This short note is aimed at providing a quantitative estimate of 
the uncertainty associated with the available models 
of the nucleon structure functions at low $Q^2$. The 
impact on the nuclear cross section is also analyzed comparing
theoretical results to the data of Refs. \cite{Sealock89,Anghinolfi96}.


The calculations of Ref. \cite{Benhar05} were carried out using the H\"ohler-Brash 
parameterization of the nucleon form factors \cite{Hohler76,Brash02}, resulting from 
a fit which includes the recent Jefferson Lab data \cite{Jones00}, and the Bodek and 
Ritchie (BR) parametrization of the proton and neutron inelastic structure functions 
\cite{Bodek81}, covering both the resonance and deep inelastic region.

The structure functions of Ref. \cite{Bodek81} have been obtained from a global fit to
the electron-proton and electron-deuteron cross sections measured at SLAC 
\cite{Bodek79}, spanning the kinematical domain $1~<~Q^2~<~20$~GeV$^2$ and 
$0.1~\lsim x~\lsim .75$, $x$ being the Bjorken scaling variable. 
As a consequence, as pointed out in Ref. \cite{Benhar05}, using the results of Bodek 
and Ritchie in the kinematical region covered by the data of
Ref. \cite{Anghinolfi96}, corresponding to $Q^2~\lsim~0.2$~GeV$^2$ at the
$\Delta$ production peak, involves an extrapolation whose validity needs to
be carefully investigated. 

Additional uncertainty is associated with the neutron structure functions $W_1^{(n)}$ 
and $W_2^{(n)}$, that have been extracted from deuteron data subtracting the proton 
contribution and unfolding nuclear effects. The authors of Ref. \cite{Bodek81} followed
the approach of Atwood and West \cite{Atwood73}, with the deuteron wave function 
obtained using the Hamada-Jonston (HJ) model of the nucleon-nucleon (NN) 
potential \cite{HJ}.

To gauge the systematic error involved in the determination of $W_1^{(n)}$ and 
$W_2^{(n)}$, we have compared the electron-deuteron cross sections 
computed using the HJ deuteron wave function to those
obtained using the state-of-the-art NN potential
of Ref. \cite{Wiringa95}, generally referred to as Argonne $v_{18}$ (A18) potential. 
The HJ and A18 models significantly differ in the description of the short 
range repulsive component of the NN force. The HJ exhibits a hard core of radius 
$r_c \sim 0.5$ fm, leading to a vanishing deuteron wave function at $r < r_c$, while
the A18 potential smoothly reaches a large but finite positive value at $r = 0$.
The differences in the short range behavior of the wave functions in coordinate space
are reflected by sizable differences in the momentum distributions $n({\bf p})$ at 
$|{\bf p}| \gsim 1$ GeV. However, as $n({\bf p})$ falls by 
a factor $\sim 10^6$ in the range $0 < |{\bf p}| < 1$ GeV, these differences have
 negligible effect on the calculated electron-deuteron cross section. For example, at 
beam energy $E_e = 4$ GeV and electron scattering angle $\theta_e=30 ^\circ$ the 
results obtained 
using the HJ and A18 momentum distributions differ by at most $2$ \% over the 
energy loss range extending from pion production threshold, corresponding to 
$\nu \sim 1.05$ GeV, to $\nu \sim 2.5$ GeV. 

We have also investigated the effect of the ambiguity implied by the relativistic 
normalization of the momentum distribution, which is obtained from the nonrelativistic
wave function in momentum space. A number of theoretical calculations have been carried 
out using the normalization of Ref. \cite{Benhar93b} instead 
of the one suggested by Attwood and West, which amounts to replacing
\beq
n({\bf p}) \rightarrow \frac{m}{\sqrt{{\bf p}^2 + m^2}}\ n({\bf p}) \ ,
\eeq
$m$ being the nucleon mass. Our results show that, as the momentum distribution is sharply peaked
at small $|{\bf p}|$, the presence of the additional normalization factor does not 
produce any appreciable effects.

Having established that the uncertainty arising from the treatment of nuclear effects 
is small, we can now address the question of whether the fit of
Ref. \cite{Bodek81} provides a reasonable description of the structure
functions at low $Q^2$. To clarify this issue we have compared the calculated 
electron-deuteron cross section, obtained using the A18 momentum distribution and the
BR structure functions, 
to the Jefferson Lab data of Refs. \cite{Niculescu99,Niculescu00} at $E_e = 2.445$ GeV and 
$\theta_e=20 ^\circ$. This kinematical setup corresponds to the lowest $Q^2$ 
at the $\Delta$ production peak available in the Jefferson Lab dataset, namely $Q^2 = 0.54$ GeV$^2$.

\begin{figure}[htb]
\centerline
{\epsfig{figure=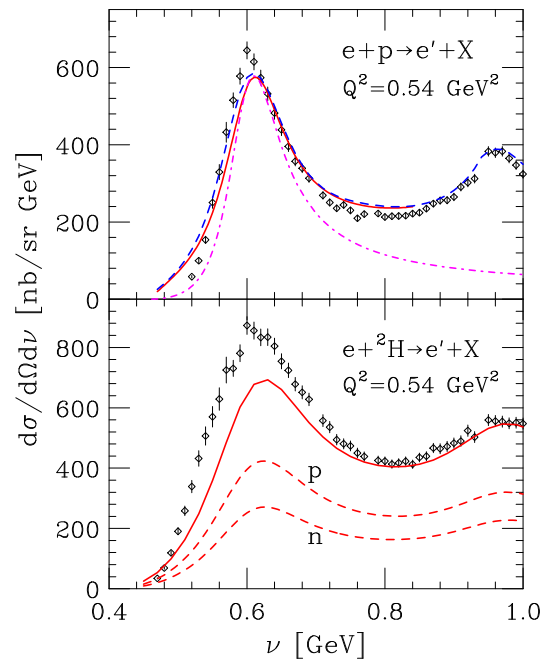,angle=000,width=7.cm}}
\caption{ (Color online) Electron-proton (upper panel) and electron-deuteron (lower panel) 
cross sections
at beam energy 2.445 GeV and scattering angle 20$^\circ$, corresponding to
Q$^2 = 0.54$ GeV$^2$ at the $\Delta$ production peak. The solid lines show the
results of calculations carried out using the structure fumctions of
Ref. \protect\cite{Bodek81}, while the dashed and dot-dash lines in the upper panel
have been obtained using the proton structure functions of
Refs. \protect\cite{Liang04,ChristyWP} and \protect\cite{Olga05a,Olga05b}, respectively.
The dashed lines labelled $n$ and $p$ in the lower panel correspond to the
neutron and proton contributions to the deuteron cross section, respectively.
The experimental data are taken from Refs. \protect\cite{Niculescu99,Niculescu00}.
}
\label{fig1}
\end{figure}

The results, represented by the solid line in the lower panel of Fig. \ref{fig1}, show 
that in the $\Delta$ 
region the measured cross section is underestimated by as much as $\sim$ 25 \%. 
On the other hand, the results displayed in the upper panel show that the BR model 
provides a much better account of the electron-proton cross section, thus implying 
that the disagreement with deuteron data is to be mostly ascribed to the neutron 
contribution.

For comparison, in the upper panel we also include the cross sections 
calculated using the structure functions resulting from a recent fit including
the Jlab data \cite{Liang04,ChristyWP} and from the dynamical model of 
Refs. \cite{Olga05a,Olga05b}. It clearly appears that the different models
describe the data in the region of the $\Delta$ peak with comparable accuracy. 
The distinctive behavior exhibited by the predictions of the model of 
Refs. \cite{Olga05a,Olga05b}, at 
both lower and higher energy loss, is to be ascribed to the fact that it does not 
include the contributions of other resonances and non resonant pion production.

In Refs. \cite{Bodek81,Bodek79} the deuteron cross section is written in the form
\beq
\sigma_d = \widetilde{\sigma}_p + \widetilde{\sigma}_n
\label{sigma_d}
\eeq
where the proton and neutron contributions, $\widetilde{\sigma}_{p,n}$ 
can in turn be expressed in terms of the proton and neutron structure functions, smeared 
by nuclear binding and Fermi motion. 
Using the Atwood-West formalism and the measured electron-proton cross 
sections, $\sigma_p$, one can then compute the ratio 
\beq
S_p = \frac{\sigma_p}{\widetilde{\sigma}_p} \ ,
\label{def:Sp}
\eeq
whose value provides a measure of nuclear effects. Note that 
Eq. (\ref{def:Sp}) is based on the premise that the ratios between the 
unsmeared and smeared structure functions, $W^p_{1,2}$ and $\widetilde{W}^p_{1,2}$, 
satisfy \cite{Bodek79}
\beq
S_p \approx \frac{W^p_1}{\widetilde{W}^p_1} \approx \frac{W^p_2}{\widetilde{W}^p_2} \ .
\eeq
The results of numerical calculations carried out in the $\Delta$ production region 
show that $S_p$, while varying sharply as a function of the 
invariant mass of the hadronic final state, $W$, exhibits a rather weak dependence 
on $Q^2$. 

\begin{figure}[htb]
\centerline
{\epsfig{figure=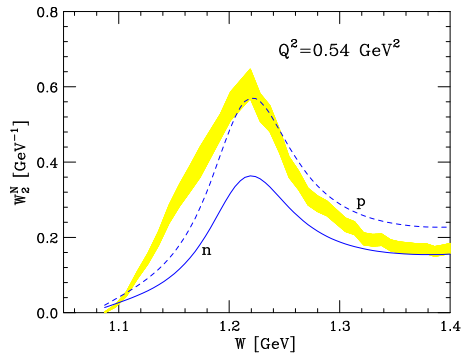,angle=000,width=6.0cm}}
\caption{ (Color online) The nucleon structure functions 
$W_2^N$ ($N=n,p$) at $E_e = 2.445$ GeV and $\theta_e = 20^\circ$,
corresponding to $Q^2=0.54$ GeV$^2$ at the $\Delta$ production peak, plotted as
a function of the invariant mass of the hadronic final state. The shaded area
represents the $W_2^n$ resulting from our analysis, while the
solid and dashed line correspond to $W_2^n$ and $W_2^p$ of Ref. \cite{Bodek81},
respectively.
}
\label{fig2}
\end{figure}

Assuming that the smearing ratios for the proton and neutron cross section, $S_p$ and $S_n$, 
be close to one another, the unsmeared neutron cross section can finally be estimated from 
\beq
\sigma_n =  S_n \widetilde{\sigma}_n \approx S_p \sigma_d - \sigma_p \ .
\eeq
The neutron structure function obtained applying the above procedure at
beam energy $E_e~=~2.445$ GeV and electron scattering angle $\theta_e = 20^\circ$, plotted
as a function of $W$, is shown by the shaded area in Fig. \ref{fig2}. 
The resulting $W_2^n$, which accounts for 
the measured deuteron cross sections by construction, turns out to be significantly 
larger than the one obtained from the fit of Ref. \cite{Bodek81}, the difference at the 
$\Delta$ production peak being $\sim 60$~\%. For comparison, the proton structure function 
$W_2^p$ is also shown.


To gauge the impact of the uncertainty in the neutron structure functions on 
the nuclear cross section, we have analyzed electron scattering off carbon 
at $E_e = 1.3$ GeV and $\theta_e~=~37.5^\circ$ and compared our results to the 
SLAC data of Ref. \cite{Sealock89}. 

\begin{figure}
\centerline
{\epsfig{figure=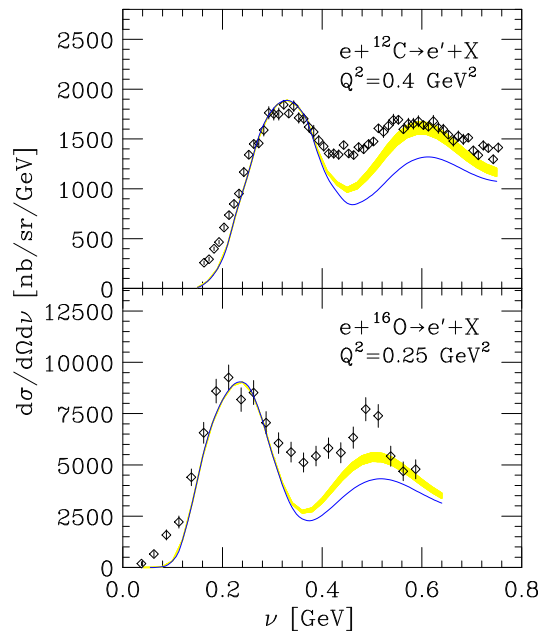,angle=000,width=7.cm}}
\caption{ (Color online) 
Upper panel: inclusive electron scattering cross section off carbon at
$E_e = 1.3$ GeV and $\theta_e = 37.5^\circ$, corresponding to $Q^2 = 0.4$
GeV$^2$ at the $\Delta$ production peak, as a function of the electron energy
loss $\nu$. The solid line corresponds to theoretical calculations carried out
using the proton and neutron structure functions of Ref. \protect\cite{Bodek81}, while
the shaded region has been obtained replacing the BR $W_1^n$ and $W_2^n$ with those
resulting from our analysis.
The data are taken from Ref. \protect\cite{Sealock89}. Lower panel: same as in the
upper panel, but
for oxygen target, $E_e = 1.2$ GeV and $\theta_e = 32^\circ$, corresponding to
$Q^2 = 0.26$ GeV$^2$ at the $\Delta$ production peak. The data are taken
from Ref. \protect\cite{Anghinolfi96}.}
\label{fig3}
\end{figure}

The shaded area in the upper panel of Fig. \ref{fig3} shows 
the results of our calculations, carried out within the Plane Wave Impulse 
Approximation (PWIA) formalism described in Ref. \cite{Benhar05} with the 
spectral function obtained from the Local Density 
Approximation (LDA) \cite{Benhar94}. We have used the proton structure functions
of Ref. \cite{Bodek81}, while for the neutron the BR fit has been 
modified according to 
\beq
W_{2}^n(Q^2,W) \rightarrow S(W) W_{2}^n(Q^2,W)\  ,
\eeq
$S(W)$ being the ratio between the results of the analysis described in the 
previous section (shaded area of Fig. \ref{fig2}) and the corresponding BR fit
(solid line of Fig. \ref{fig2}). In view of the fact that in the kinematical 
setup of the carbon data the value of $Q^2$ at the $\Delta$ production peak
is $\sim$ 0.4 GeV$^2$, using $S(W)$ extracted at 
$Q^2 \sim 0.5$ GeV$^2$ appears to be reasonable.

Comparison with the solid line, obtained using the BR fit for both the proton and
the neutron, shows that the neutron structure functions that reproduce the deuteron data 
at $Q^2$ = 0.54 GeV$^2$ also provide a much better description of carbon data at 
electron energy loss $\nu \gsim 0.6$ GeV. Discrepancies between theory and
experiment still persist at lower $\nu$, in the region of the dip between the 
quasi-elastic and the $\Delta$ production peak.

The inclusive cross section in the dip region is known to be affected by
processes in which the electron couples to the two-body component of the nuclear
electromagnetic current \cite{Carlson98}, arising from $\pi$ and $\rho$
meson exchange between nucleons. However, the amount of strength associated
with meson exchange currents (MEC) is still somewhat controversial, and
the possibility that the cross section in the dip region might be ascribed
to an asymmetry in the shape of the quasi-elastic peak has also been
suggested (see, e.g., \cite{Amaro05} and references therein).
The contribution of MEC is not taken into account in the PWIA approach
of Ref. \cite{Benhar05}, and its significance in the kinematical regime under
discussion should be carefully investigated.

The lower panel of Fig. \ref{fig3} shows a comparison between our theoretical results 
and the oxygen data discussed in Ref. \cite{Benhar05}. It clearly appears that at 
the low $Q^2$ of the data of Ref. \cite{Anghinolfi96} the enhancement 
resulting from using the neutron structure functions obtained from our analysis is 
not sufficient to explain measured cross section. 

The results discussed in this paper show that the structure 
functions obtained from the global fit of Refs. \cite{Bodek79,Bodek81}, while
providing a quantitative account of proton data, fail to explain the measured 
deuteron cross sections at low Q$^2$ ($< 1$ GeV$^2$). 

Employing a simple and somewhat crude procedure, similar to the one followed by 
Bodek and Ritchie, we have extracted the neutron structure functions at 
$Q^2 \sim 0.5$ GeV$^2$ from the Jefferson Lab electron-proton and electron-deuteron data. 
The resulting $W_1^n$ and $W_2^n$  turn out to be much larger than those of 
Ref. \cite{Bodek81} in the region of the $\Delta$ production peak.

When used to calculate the nuclear cross section in the impulse approximation scheme,
the neutron structure functions obtained from our analysis provide a satisfactory
description of carbon data at Q$^2=0.4$ GeV$^2$. On the other hand, oxygen data
at lower Q$^2$ are still sizably underestimated. This feature is likely to reflect 
an appreciable Q$^2$-dependence of the function $S(W)$, describing the difference 
between our results and the BR fit. 

The results of Figs.\ref{fig1} and \ref{fig3} have been obtained neglecting final state 
interactions (FSI), which are known to lead to a redistribution of the
inclusive strength \cite{Benhar93a}. Within the approach of
Ref. \cite{Benhar93b}, FSI effects in the quasi-elastic channel are 
included through a folding procedure that accounts for 
nucleon-nucleon rescattering processes. However, the extension to 
inelastic channels involves additional problems and has not been carried 
out yet. 
To estimate the relevance of FSI in the $\Delta$ production region we have 
folded the PWIA inelastic cross sections of Figs. \ref{fig1} and \ref{fig3} using the 
folding functions obtained from the approach of Ref. \cite{Benhar93b}.
The results, showing that the main effect is a quenching of the peak 
of less than $\sim 2$\% and $4$\% in deuteron and carbon, respectively, suggest that 
FSI do not significantly affect our analysis.

The ability of dynamical models \cite{Olga05a,Olga05b,Sato96,Sato06} to explain
the nuclear cross section should also be investigated. The results of a recent
calculation \cite{Sato06} based on the model of Ref. \cite{Sato96},
are in fairly good agreement with the data of Ref. \cite{Sealock89}
in the $\Delta$ production region at $Q^2 \sim$ 0.2 GeV$^2$. However, the
fact that measured quasi-elastic cross section is sizably overestimated
seems to point to deficiencies in the treatment of nuclear effects.

In conclusion, our study suggests that the neutron structure functions in the $\Delta$ 
production region at low $Q^2$, which is still poorly known, may be extracted from a fit 
to the upcoming deuteron data from Jefferson Lab \cite{Juppiter-NUINT05}. Valuable 
complementary information will also come from the direct measurement of the free neutron 
structure functions at $1 < Q^2 < 5$, presently being carried out by 
the Jefferson Lab E03-12 (BoNuS) collaboration \cite{Bonus}.




\begin{thebibliography}{99}



\bibitem{NUINT01}
Proceedings of NuInt01, Eds. J.G. Morfin, M. Sakuda and
Y. Suzuki. Nucl. Phys. B (Proc. Suppl.) {\bf 112} (2002).

\bibitem{NUINT04}
Proceedings of NuInt04, Eds. F. Cavanna, P. Lipari, C. Keppel
 and M. Sakuda. Nucl. Phys. B (Proc. Suppl.) {\bf 139} (2005).

\bibitem{Benhar06}
O. Benhar, D. Day and I. Sick, nucl-ex/0603029. Submitted for publication
in Reviews of Modern Physics.


\bibitem{Benhar93a}
O. Benhar, V.R. Pandharipande and S.C. Pieper. 
Rev. Mod. Phys. {\bf 65} (1993) 817.

\bibitem{Benhar05}
O. Benhar, N. Farina, H. Nakamura, M. Sakuda and R. Seki,
Phys. Rev. D {\bf 72} (2005) 053005.

\bibitem{Sealock89}
R.M. Sealock {\it et al}, Phys. Rev. Lett. {\bf 62} (1989) 1350.

\bibitem{Anghinolfi96}
M. Anghinolfi {\it et al.}, Nucl. Phys. A {\bf 602} (1996) 405.

\bibitem{Hohler76} G. H\"ohler {\it et al.},
Nucl.\ Phys. {\bf B114} (1976) 505.

\bibitem{Brash02} E.J. Brash, A. Kozlov, Sh. Li, and G.M. Huber,
Phys. Rev. C {\bf 65}, 051001(R) (2002) 051001(R).

\bibitem{Jones00} M.K. Jones {\it et al.},
Phys. Rev. Lett. {\bf 84} (2000) 1398.

\bibitem{Bodek81}
A. Bodek and J.L. Ritchie, Phys. Rev. D {\bf 23} (1981) 1070.

\bibitem{Bodek79}
A. Bodek {\it et al.}, Phys. Rev. D {\bf 20} (1979) 1471.

\bibitem{Atwood73}
W.B. Atwood and G.B. West,  Phys. Rev. D {\bf 7} (1973) 773.

\bibitem{HJ}
T. Hamada and I.D. Johnston, Nucl. Phys. A {\bf 34} (1982) 382.
 
\bibitem{Wiringa95}
R.B. Wiringa, V.G.J. Stoks and R. Schiavilla,
Phys. Rev. C {\bf 51} (1995) 38.

\bibitem{Benhar93b}
O. Benhar and V.R. Pandharipande, Phys Rev. C {\bf 47} (1993) 2218.


\bibitem{Niculescu99}
I. Niculescu, PhD Thesis, Hampton University, 1999. Unpublished.
  
\bibitem{Niculescu00}
I. Niculescu {\it et al}, Phys. Rev. Lett. {\bf 85} (2000) 1186.

\bibitem{Liang04}
Y. Liang, M.E. Christy, R. Ent and C.E. Keppel (Jefferson Lab Hall C 
E94-110 Collaboration), nucl-ex/0410027.

\bibitem{ChristyWP}
http://www.jlab.org/\~{}christy/cs\_fits/cs\_fits.html

\bibitem{Olga05a}
O. Lalakulich and E.A. Paschos, Phys. Rev. D {\bf 71} (2005) 074003.

\bibitem{Olga05b}
O. Lalakulich, E.A. Paschos and G. Piranishvili,  Nucl. Phys. B
(Proc. Suppl.) \bf{159} (2006) 133.

Nuclear Physics B (Proc. Suppl.).


\bibitem{Benhar94}
O. Benhar, A. Fabrocini, S. Fantoni and I. Sick, 
Nucl. Phys. A579 (1994) 493.

\bibitem{Carlson98}
J. Carlson and R.Schiavilla, Rev. Mod. Phys. {\bf 70} (1998) 743.

\bibitem{Amaro05}
J.E. Amaro, M.B. Barbaro, J.A. Caballero, T.W. Donnelly, A. Molinari and
I. Sick, Phys. Rev. C {\bf 71} (2005) 015501.

\bibitem{Sato96}
T. Sato and T.-S.H. Lee, Phys. Rev. C {\bf 54} (1996) 2660.

\bibitem{Sato06}
T. Sato, B. Szczerbinska, K. Kubodera and T.-S.H. Lee, Nucl. Phys. B 
(Proc. Suppl.) \bf{159} (2006) 141.

\bibitem{Juppiter-NUINT05}
V. Tvaskis, J. Steinman and J. Bradford,  Nucl. Phys. B
(Proc. Suppl.) \bf{159} (2006) 163.

\bibitem{Bonus}
www.jlab.org/\~{}kuhn/BoNuS\_Welcome.html

\end{thebibliography}
\end{document}